\documentclass[11pt,a4paper]{article}
\usepackage{acl2015}
\usepackage{times}
\usepackage{url}
\usepackage{latexsym}
\usepackage{multirow}
\usepackage{amssymb,amsmath,graphicx,amsfonts,epsfig,epstopdf,datetime, mathrsfs,algorithm,algorithmic}
\usepackage{float}
\usepackage{stfloats}

\title{Stochastic Top-k ListNet}

\author{Tianyi Luo$^1$, Dong Wang$^{1,2}$, Rong Liu$^{1,3}$, Yiqiao Pan$^{1,4}$ \\
  $^1$CSLT, RIIT, Tsinghua University, China \\
  $^2$Tsinghua National Lab for Information Science and Technology\\
  $^3$Huilan Limited, Beijing, China \\
  $^4$Beijing University of Posts and Telecommunications, China \\
  {\tt \{lty, lr, pyq\}@cslt.riit.tsinghua.edu.cn} \\
  {\tt wangdong99@mails.tsinghua.edu.cn}
  \\}

\date{}

\begin{document}
\maketitle
\begin{abstract}

ListNet is a well-known listwise learning to rank model and has gained much attention in recent years. A particular problem of ListNet, however, is the high computation complexity in model training, mainly due to the large number of object permutations involved in computing the gradients. This paper proposes a stochastic ListNet approach which computes the gradient within a bounded permutation subset. It significantly reduces the computation complexity of model training and allows extension to Top-k models, which is impossible with the conventional implementation based on full-set permutations. Meanwhile, the new approach utilizes partial ranking information of human labels, which helps improve model quality. Our experiments demonstrated that the stochastic ListNet method indeed leads to better ranking performance and speeds up the model training remarkably.

\end{abstract}

\section{Introduction}
\label{sec:intro}

Learning to rank aims to learn a model to re-rank a list of objects, e.g., candidate documents in document retrieval. Recent studies show that listwise learning delivers better performance in general than traditional pairwise learning~\cite{liu2009learning}, partly attributed to its capability of learning human-labelled scores as a full rank list. A potential disadvantage of listwise learning, however, is the high computation complexity in model training, which is mainly caused by the large number of permutations of the objects to rank.

A typical listwise learning method is the ListNet model proposed by~\newcite{cao2007learning}. This model has been utilized to tackle many ranking problems, e.g. modeling the hiring behavior in online labor markets~\cite{kokkodis2015hiring}, ranking sentences in document summarization~\cite{jin2010comparative}, improving detection of musical concepts~\cite{yang2009improving} and ranking the results in video search~\cite{yang2008video}. Basically, ListNet implements the rank function as a neural network (NN), with the objective function set to be the cross entropy between two probability distributions over the object permutations, one derived from the human-labelled scores and the other derived from the model prediction (network output). In order to deal with the high computation complexity associated with the large number of permutations, \newcite{cao2007learning} proposed a Top-k approach, which clusters the permutations by the first $k$ objects, so the number of distinct probabilities that need to evaluate in model training reduces from $n!$ to $\frac{n!}{(n-k)!}$, where $n$ is the number of objects in the list.

To ensure efficiency, $k=1$ was selected in the seminal paper~\cite{cao2007learning} and in the open source implementation of RankLib~\cite{dang2013ranklib}. This Top-1 approach is a harsh approximation to the full listwise learning and may constrain the power of the ListNet method.
We therefore seek to extend the Top-1 approximation to Top-k ( k $>$ 1) models.

The major obstacle for the Top-k extension is the large number of permutations, or more precisely, permutation classes in the Top-k setting. A key idea of this paper is that the rank information involved in the permutation classes is highly redundant and so a small number such permutation classes are sufficient to convey the rank information required to train the model. Meanwhile, the partial rank information associated with the subset of permutation classes may
represent more detailed knowledge for model training, leading to better ListNet models.

Based on these two conjectures, we propose a stochastic ListNet method, which samples a subset of the permutation classes (object lists) in model training
and based on this subset to train the ListNet model.  Three methods are proposed to conduct the sampling. In the uniform distribution method, the candidate objects are selected following a uniform distribution; in the fixed distribution method, the candidate objects are selected following a distribution derived from the human-labeled scores; in the adaptive distribution method, the candidates are selected following a distribution defined by the rank function, i.e., the neural network output. Experimental results demonstrated that the stochastic ListNet method can significantly reduce the computation cost in model training. In fact, if the size of the permutation subset is fixed, the computation complexity is bounded, which allows training Top-k models where $k$ is large. Meanwhile, better performance was obtained with the stochastic ListNet approach, probably due to the learning of partial rank information.

The contributions of the paper are three-fold: (1) proposes a stochastic ListNet method that significantly reduces the training complexity and delivers better ranking performance; (2) investigates Top-k models based on the stochastic ListNet, and studies the impact of a large $k$; (3) provides an open source
implementation based on RankLib.

The rest of the paper is organized as follows. Section~\ref{sec:rel} introduces some related works, and Section~\ref{sec:method} presents the stochastic ListNet method. Section~\ref{sec:exp} presents the experiments, and the paper is concluded by Section~\ref{sec:con}.

\section{Related Work}
\label{sec:rel}


This work is an extension of the Top-k ListNet method proposed by~\newcite{cao2007learning}. The novelty is that we propose a stochastic learing method which not only speeds up the model training but also produces stronger models. The code is based on the Top-1 ListNet implementation of RankLib~\cite{dang2013ranklib}.

Another related work is the SVM-based pairwise learning to rank model based on stochastic gradient descent (SGD)~\cite{sculley2009large}. In this approach,
training instances (queries) are selected randomly and for each query, a number of object pairs are sampled from the object list.
These pairs are used to train the SVM model. In the stochastic ListNet method proposed in this paper, the randomly selected training samples are
permutation classes (object lists) rather than pairs of objects, and a set of object lists rather than a single pair forms a training sample.



\section{Methods}
\label{sec:method}

\subsection{Review of ListNet}

The ListNet approach proposed by~\newcite{cao2007learning} trains a neural network which predicts the scores $z^{(i)}$ of a list of candidate objects $x^{(i)}$ given a query $q^{(i)}$, formulated by $z^{(i)} = f_w(x^{(i)})$, where $f_w$ stands for the scoring function defined by the NN. The objective function is given by:
\begin{eqnarray}
\nonumber
   \mathcal{L} &=& \sum_i \mathcal{L}({y}^{(i)},{z}^{(i)}) \\
\label{eq:cost}
               &=& \sum_i \sum_{\forall g \in \mathscr{G}_k}P_{y^{(i)}}(g)log(P_{z^{(i)}}(g))
\end{eqnarray}
\noindent where $y^{(i)}$ denotes the human-labelled scores, and $\mathscr{G}_k$ is the set of permutation classes defined by:
\begin{eqnarray}
\nonumber
\mathscr{G}_k &=& \{\mathscr{G}_k(j_1,j_2,...,j_k) | j_t = 1,2,...,n, \\
&& \ s.t. \ j_u \ne j_v \ \ \ for \ \ \forall u \ne v\}
\label{eq:g}
\end{eqnarray}

\noindent where $n$ is the number of candidate objects, $j_t$ is the object ranked at the $t$-th position, and $\mathscr{G}_k(j_1,j_2,...,j_k)$ is a permutation class which involves all the permutations whose first $k$ objects are exactly $(j_1,j_2,...,j_k)$. Following~\newcite{cao2007learning}, the probability of $\mathscr{G}_k(j_1,j_2,...,j_k)$ can be computed by:
\begin{equation}
\label{eq:pg}
P_{s}(\mathscr{G}(j_1,j_2,...,j_k)) = \prod_{t=1}^{k} \frac{e^{s_{j_{t}}}} {\sum_{l=t}^{n} e^{s_{j_l}}}.
\end{equation}
\noindent where $s_{j_{t}} $ is the score of object at position $j_t(t = 1, 2,,, k)$ at a certain permutation. By this definition of permutation probability, Eq.~(\ref{eq:cost}) defines a cross entropy between the distributions over permutations (precisely, permutation classes) derived from the human-labelled scores and the NN-predicted scores. Therefore, optimizing the objective function Eq.~(\ref{eq:cost}) with respect to the NN model $f_w$ leads to a scoring function that approximates the human-labelled ranking.

\subsection{Stochastic Top-k ListNet}

A particular difficulty of the Top-k ListNet method is that it requires  very demanding computation in model training.
Refer to Eq. (\ref{eq:g}), the permutation set $\mathscr{G}_k$ involves $\frac{n!}{(n-k)!}$ members,
and for each member, computing its probability involves $\frac{(2n-k+1)k}{2}$ summations plus $k$
multiplications and divisions. To let the algorithm practical, $k$=$1$ was selected in~\cite{cao2007learning},
as well as the public toolkit RankLib~\cite{dang2013ranklib}. Although this is a good solution and reduces
computation dramatically, we argue that this approach largely buries the power of ListNet. In fact,
setting $k$=$1$ effectively marginalizes all the probabilities over the candidate objects of a
permutation class except the top one. By this approximation, Eq.~(\ref{eq:pg}) reduces to a softmax over the
candidate objects, which means that it actually focuses on how the probabilities are distributed
over \emph{individual} objects, rather than how the probabilities are distributed
over \emph{object lists}. This potentially loses much rank information involved in the
human labels.

Another disadvantage of the Top-1 model is that it learns the rank information
of the \emph{full list}, but ignores the rank information of \emph{partial sequences},
which may lead to ineffective learning.
As an example, considering an object list where the score of the most relevant object is much higher
than the scores of others, then the learning is dominated by the highest score, and largely throws
away the rank information conveyed by the scores of other objects. It would be quite helpful if
the rank information involved in partial sequences of the candidate objects can be learned. Top-k
models place distributions over object lists (in length $k$), and so can learn partial
sequences of objects.

We are interested in how to learn Top-k ($k$ $>$ 1) models while keeping the computation tractable.
To achieve the goal, we propose a stochastic ListNet approach, which
samples a small set of the Top-k permutation classes (object lists), and train the Top-k model based on this
small set instead of the full set of permutation classes. As a comparison, the full set of
permutation classes of the Top-k model is $\frac{n!}{(n-k)!}$, which is
computationally prohibitive if $k$ $>$ 1. With stochastic ListNet, a subset of the permutation classes
that involves only $l$ members are randomly selected. Training the Top-k model based on
this subset greatly reduces the computation cost, even with a large $k$. In fact, the subset approach
imposes a bound of the computation cost that is largely
determined by the the size of the subset ($l$), while independent of the total number of objects
$n$ and the model order $k$.

Interestingly, the stochastic approach offers not only quick learning, but also a chance of learning
partial ranks. This is obvious because only a subset of the object lists are selected in model training,
and so the rank information involved in the subset of the permutation classes can be learned. With the Top-1
model, partial ranks reduces to partial sequences since each object list involves only one object. As we
have discussed, learning partial sequences is an advantage of Top-k models with $k > 1$. This means that
stochastic Top-1 ListNet possesses some advantages of Top-k ListNet, while the computation cost is much lower.

\subsection{Sampling methods for stochastic ListNet}

The training process of stochastic ListNet starts from sampling $l$ permutation classes, or object lists. For each object list,
$k$ objects are sampled following a particular distribution.
As mentioned in Section~\ref{sec:intro}, three distributions are studied in this paper:  uniform
distribution, fixed distribution and adaptive distribution. They are presented as follows.

{\bf Uniform distribution sampling}:
In this method, all the $k$ objects of a particular object list are sampled with an equal probability. This sampling method is simple but biased
towards irrelevant candidates, since there are much more irrelevant objects than relevant ones in the training data. A re-sampling
approach is proposed to remedy the bias, as will be discussed in Section~\ref{sec:exp}.

{\bf Fixed distribution sampling}:
In this method, the objects are sampled following a distribution proportional to the human-labelled scores. For instance, in the LETOR
dataset that is used in this study, each candidate object (document) is labelled as 2 (very relevant), 1 (relevant) or 0 (irrelevant). These scores are normalized by
softmax and are used as the probability distribution when sampling objects. Because the probabilities of relevant objects are larger than those of irrelevant
objects, more relevant objects would be selected by this sampling approach in model training.

{\bf Adaptive distribution sampling}:
The fixed distribution sampling mentioned above relies on human-labelled scores, which may be impacted by label errors. Moreover, the absolute
values of human labels are not good measures of object relevance. To solve these problems, we choose the outputs of the `current' neural
network as the relevance scores, and sample the objects according to these scores. Note that the network
outputs are natural measures of object relevance based on the present ranking model. As the model (the neural network) keeps updated during
model training, the relevance scores are accordingly changed. In each iteration, the relevance scores are
re-calculated, and the sampling is based on the new scores in the next iteration.

\subsection{Gradients with linear networks}

\newcite{cao2007learning} optimized the ListNet model by gradient descent. For each query, the learn rule is formuated by:
\[
 w = w - \eta \Delta w
\]
\noindent where $\eta$ is the learning rate, and $w$ denotes the parameters of the model $f_w$. $\Delta w$ denotes the gradient and it can be computed as follows:
\[
\Delta w = \sum_{\forall g \in \mathscr{G}_k} \frac{\partial P_{z^{(i)}(f_w)}(g)}{\partial w}  \frac{P_{y^{(i)}}(g)}{P_{z^{(i)}(f_w)}(g)}.
\]
\noindent For simplicity, a linear NN model was used by~\newcite{cao2007learning}. This has been adopted in our study as well, written by $z^{(i)} = f_w (x^{(i)}_j) = w^T x^{(i)}_j$, where $x^{(i)}_j$ denotes the feature vector of the $j$-th object of the $i$-th query. In the case of the Top-1 model, it shows that:
\begin{eqnarray}
\nonumber
\Delta w= \sum_j [\sigma(z^{(i)},j) - \sigma(y^{(i)},j)] x^{(i)}_j
\end{eqnarray}
\noindent where $\sigma(s,j)$ is the $j$-th value of the softmax function of the score vector $s$, given by:
\[
\sigma(s^{(i)},j) = \frac{e^{s^{(i)}_j}} {\sum_{t=1}^{n^{(i)}} e^{s^{(i)}_t} }.
\]
In the case of the Top-k model, the gradient(Derivative of cross entropy between $P_{z^{(i)}}$ and $P_{y^{(i)}}$ when $k$ $>=$ 2) is a bit complex, but still manageable:

\begin{equation}
\begin{aligned}
\label{eq:topk}
& \Delta w = \sum_{g \in \mathscr{G}_k} [(\prod_{t=1}^{k} \hat{\sigma}(y^{(i)}, t))\cdot \\
& (\sum_{f=1}^{k} \{ x^{(i)}_{j_f} - \sum_{v=f}^{n^{(i)}} \hat{\sigma}(z^{(i)}, v) x^{(i)}_{j_v} \})]
\end{aligned}
\end{equation}

\noindent where $\hat{\sigma}(\cdot)$ defines a `partial' softmax(The ¡®partial softmax¡¯ means that the $\sigma(s,f)$ has a similar form as softmax, however when computing the value for each f, the denominator is not the summation from 1 to n, instead a ¡®partial sequence¡¯ from f to n.), given by:
\[
\hat{\sigma}(s^{(i)}, f) = \frac{e^{s^{(i)}_{j_f}}} {\sum_{t=f}^{n^{(i)}} e^{s^{(i)}_{j_t}}}.
\]

\subsection{Stochastic Top-k ListNet algorithm}

We present the stochastic Top-k ListNet algorithm, by employing the techniques described above. The gradient descent (GD) approach is adopted. All the training samples are processed sequentially in an iteration. The training runs several iterations until the convergence criterion is reach. Another
detail is that the learning rate is multiplied by $0.1$ whenever the objective function is worse than the previous iteration. The procedure is illustrated in Algorithm~\ref{alg:listnet}, where $\mathcal{L}(t)$ denotes value of the objective function after the $t$-th iteration.

\begin{algorithm}[h]
    \caption{Stochastic Top-k ListNet}
    \label{alg:listnet}
    \begin{algorithmic}[1]
        \REQUIRE ~~
            \\
            Input:\\
            $\mathscr{D} = \{(q^{(1)}, x^{(1)}, y^{(1)}), ..., (q^{(m)}, x^{(m)}, y^{(m)}) \}$: training data\\
            T: number of iterations \\
            $\eta$: learning rate \\
        \ENSURE ~~
        \STATE Randomly initialize $w$
        \FOR{$t=1$ to T}
                \FOR{$i=1$ to m}
                    \STATE select the $i$-th training instance $(q^{(i)},x^{(i)},y^{(i)}) \in \mathscr{D}$
                    \STATE Sample the permutation classes $\mathscr{G}_k $
                    \STATE Compute $\Delta w$ according to Eq. (\ref{eq:topk})
                    \STATE Update $f_w$: $w = w - \eta \Delta w$
                \ENDFOR
                \IF {$\mathcal{L}(t) < \mathcal{L}(t-1)$}
                    \STATE $\eta = 0.1 \eta$
                \ENDIF
        \ENDFOR
    \end{algorithmic}
\end{algorithm}

\section{Experiments}
\label{sec:exp}

\subsection{Data}

The proposed stochastic Top-k ListNet method is tested on the document retrieval task based on the
MQ2008 dataset of LETOR 4.0~\cite{liu2007letor}. This database was released in early 2007 and has been widely used in learning to
rank studies. It contains queries and corresponding candidate documents. The human-labelled scores
are among three values $\{0, 1, 2\}$, representing little, medium, and strong relevance
between queries and candidate documents, respectively. The training set, validation set and test data all
contain $784$ queries. The document features used in this study include term frequency, inverse document frequency, BM25,
and language model scores for IR. Some new features proposed recently are also included, such as
HostRank, feature propagation, and topical PageRank.

\subsection{Experiment Setup}

In our experiments, we consider Top-k models where k = 1, 2, 3, and 4.
Although any $k$ is possible with the proposed stochastic ListNet, we will show that simply increasing the
model order $k$  does not improve performance. The P@1 and P@10 performance is used as the evaluation metric.

Specially, for all the three distribution sampling methods, the sampling process involves two steps: pre-selection
and re-sampling. The pre-selection step samples a list of documents
following three distributions mentioned above, and in the re-sampling step, document lists including more
relevant documents are retained with
a higher probability. For example, denoting the pre-selected document list by ($v_1$,$v_2$,...,$v_k$)
where $k$ is the length of the list, and denoting the corresponding human-labelled
scores by ($s_1$, $s_2$,...,$s_k$), the probability that the list is retained is given by

\[
\frac{\sum_{i=1}^{k} s_i}{kS}
\]
\noindent where $S$ is the maximum value of the human-labelled scores, which is $2$ in our case.
The re-sampling approach is designed to encourage document lists containing more relevant documents,
which is the most important for the uniform distribution sampling.

In stochastic Top-k ListNet, the learning rate is set as $10^{-3}$ for $k$ = 1, and
$10^{-5}$ for $k$ $>$ 1.  These values are set to achieve the best
performance on the validation set. Another important parameter of the stochastic Top-k ListNet
approach is the number of samples of the document lists (or the size of subset of permutation classes selected),
denoted by $l$. Various settings of $l$ are experimented with in this study. To eliminate randomness
in the results, all the experiments are repeated $20$ times and the averaged performance is reported.

\subsection{Experimental results}

\begin{figure}[!htb]
\centering
\epsfig{figure=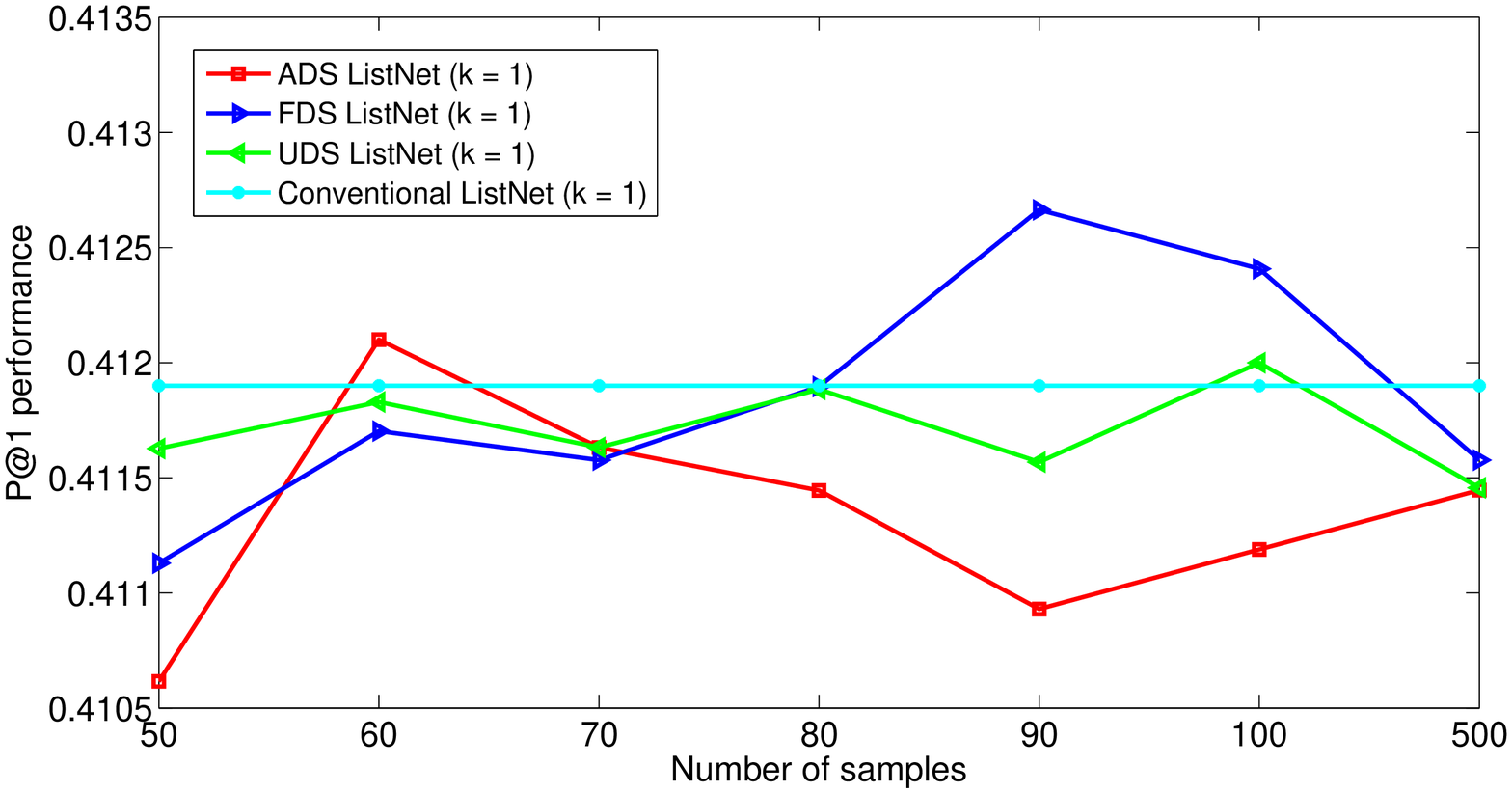,width=8cm}
\caption{The P@1 performance on the test data with the Top-1 ListNet utilizing the three sampling approaches. The size of the permutation subset varies from 50 to 500. }
\label{fig:k1}
\end{figure}

\begin{figure}[!htb]
\centering
\epsfig{figure=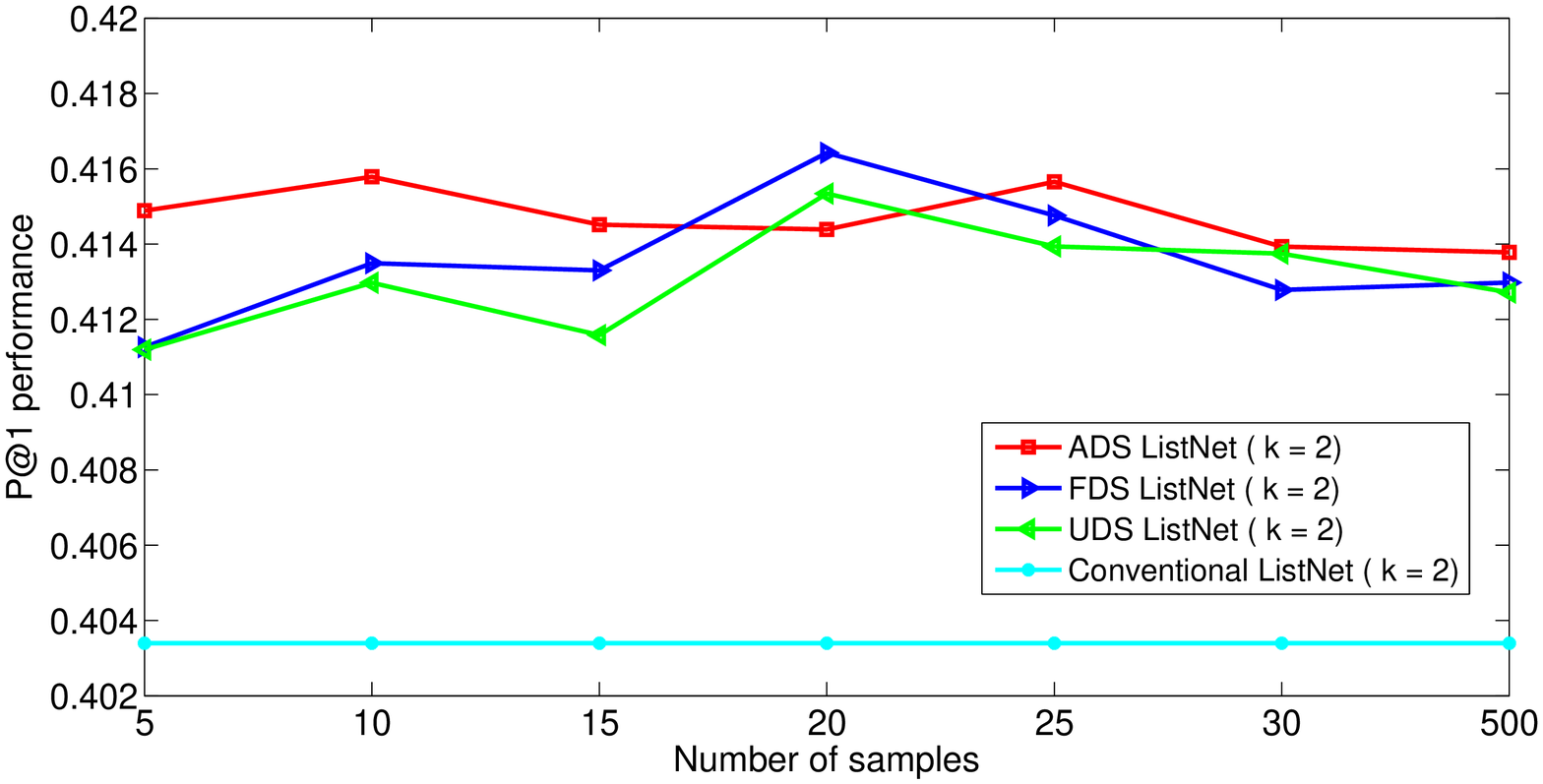,width=8cm}
\caption{The P@1 performance on the test data with the Top-2 ListNet utilizing the three sampling approaches. The size of the permutation subset varies from 5 to 500. }
\label{fig:k2}
\end{figure}

\begin{figure}[!htb]
\centering
\epsfig{figure=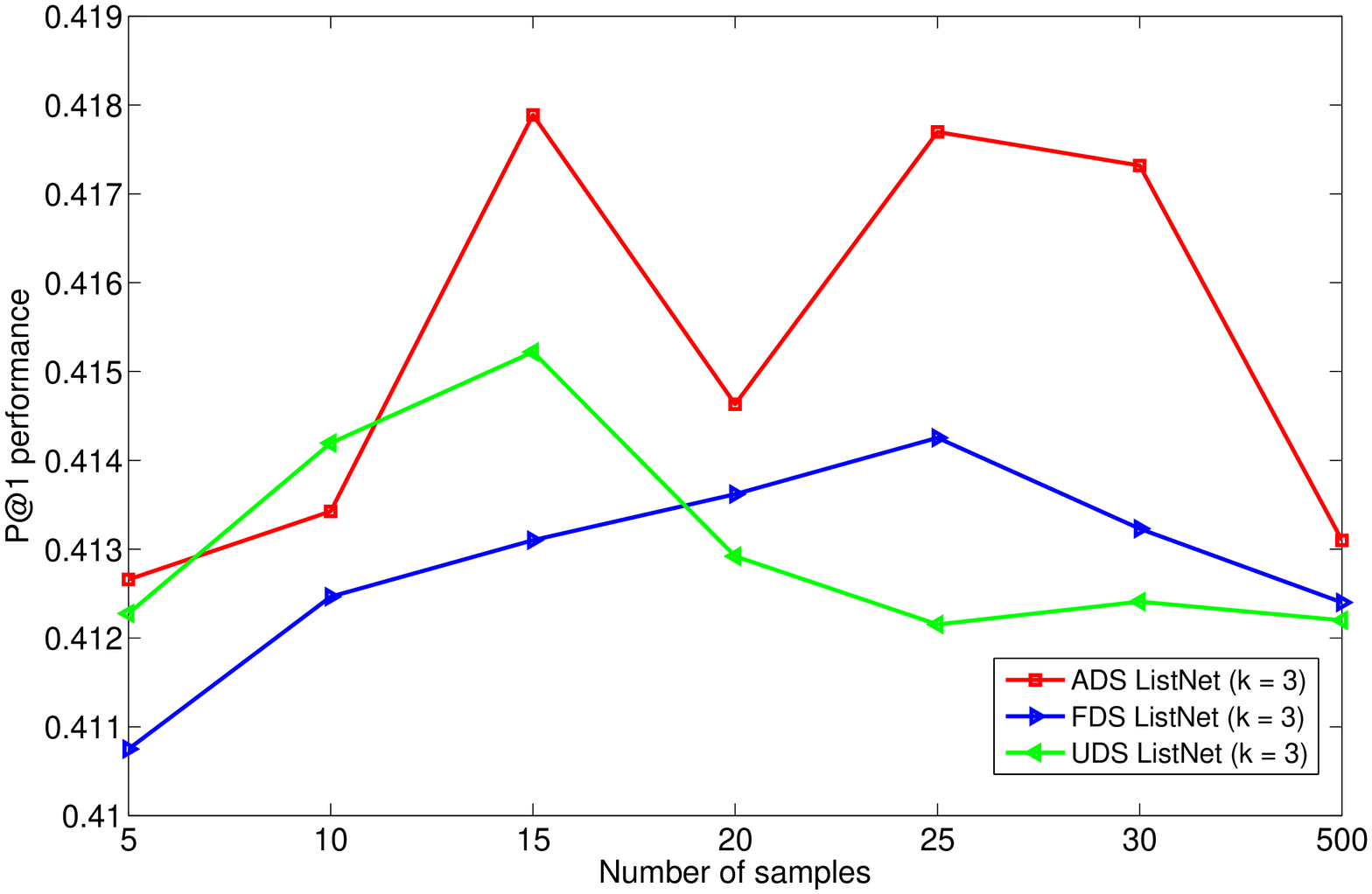,width=8cm}
\caption{The P@1 performance on the test data with the Top-3 ListNet utilizing the three sampling approaches. The size of the permutation subset varies from 5 to 500. }
\label{fig:k3}
\end{figure}

\begin{figure}[!htb]
\centering
\epsfig{figure=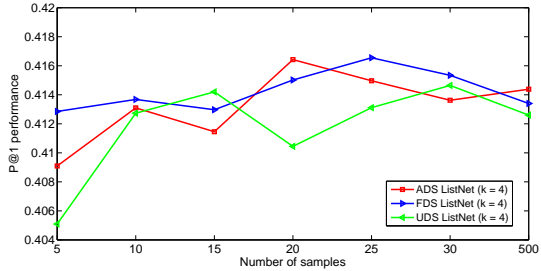,width=8cm}
\caption{The P@1 performance on the test data with the Top-4 ListNet utilizing the three sampling approaches. The size of the permutation subset varies from 5 to 500. }
\label{fig:k4}
\end{figure}

The P@1 results on the test dataset with different orders of Top-k ListNet are reported in Figure~\ref{fig:k1} to Figure~\ref{fig:k4}.
In each figure, the number of document lists varies from $5$ to $500$. For comparison, the results with the conventional
ListNet are also presented. Note that the re-sampling approach was not applied to the Top-1 model as we found it caused performance
reduction. This is perhaps because the sampling space is small with the Top-1 model, and so re-sampling tends to cause
over-emphasis on relevant documents.

From these results, we first observe that stochastic ListNet with either fixed or adaptive distribution
sampling tends to outperform the conventional ListNet approach, particularly with a large $k$. This confirms our argument
that rank information can be learned from a subset of the permutation classes that are randomly selected, and the partial rank learning
can lead to even better performance than the full rank learning, the case of conventional ListNet. This is an interesting result and
demonstrates the stochastic ListNet is both faster and better than the conventional ListNet. It is also seen
that the adaptive distribution sampling performs slightly better than the fixed distribution sampling.
This is not surprising as the adaptive distribution sampling uses a more reasonable relevance score (neural network output)
to balance relevant and irrelevant documents. The uniform distribution sampling performs a little worse than the other two sampling methods, probably
caused by the less informative uniform distribution.

Another observation is that in all the four figures, the performance of the stochastic ListNet
methods increases with more samples of the object lists. However if there are too many samples,
the performance starts to decrease. This can be explained by the fact that the sampling prefers
relevant documents which are more informative.
A larger sample set often includes more informative documents; however if the set is too large,
many irrelevant documents will be selected and the performance is reduced. In the case that
the number of samples is very large ($500$ for example for Top-1), the stochastic ListNet falls back to the conventional ListNet,
and their performance becomes similar.

Comparing the results with different $k$, it can be seen that a larger $k$ leads to a better performance
with stochastic ListNet. This confirms that high-order Top-k models can learn more ranking information. However,
this is not necessarily the case with the conventional ListNet. For example, the Top-2 model does not offer
better performance than the Top-1 model. This is perhaps because high-order Top-k models consider a large number
of document lists and most of them are not informative, which leads to ineffective learning. Remind that
the conventional ListNet is a special case of the stochastic ListNet with a very large sample set,
and we have discussed that an over large sample set actually reduces performance.

\begin{table*}[bp]
\begin{center}
\begin{tabular}{|c|c|c|c|c|c|c|c|c|c|}
\hline
      &         &          &      &\multicolumn{3}{c|}{P@1} &\multicolumn{3}{c|}{P@10}\\
\hline
Model &  Top-k & Sampling          &  Time (s) &  Train & Val. & Test &  Train & Val. & Test\\
\hline
C-ListNet           & k=1  &  -        &2.509   &0.4101 &0.4107 &0.4119 & \bf{0.2684} & \bf{0.2684} & 0.2676\\
S-ListNet         & k=1  & UDS   &0.753    & 0.4097 & \bf{0.4106} & 0.4120 &0.2680 &0.2683 &0.2676\\
S-ListNet         & k=1  & FDS     &0.391    & 0.4094 & 0.4090 & \bf{0.4127} &0.2679 &0.2681 &0.2676\\
S-ListNet         & k=1  & ADS  & \bf{0.375}    & \bf{0.4102} & 0.4097 & 0.4121 &0.2680 &0.2682 &\bf{0.2677}\\
\hline
C-ListNet           & k=2  & -         &2275.5    & 0.4119 & 0.4043 & 0.4043 &0.2678 &0.2674 &0.2674\\
S-ListNet         & k=2  & UDS   &2.898    & 0.4140 & 0.4143 & 0.4130 &0.2682 &0.2686 &0.2681\\
S-ListNet         & k=2  & FDS     &2.410    & 0.4145 & 0.4144 & \bf{0.4164} &0.2684 &0.2688 &0.2684\\
S-ListNet         & k=2  & ADS  & \bf{2.013}    & \bf{0.4162} & \bf{0.4168} & 0.4145 &\bf{0.2686} &\bf{0.2689} &\bf{0.2687}\\
\hline
S-ListNet         & k=3  & UDS   &4.358    & 0.4167 & 0.4204 & 0.4152 &0.2686 &0.2681 &0.2680\\
S-ListNet         & k=3  & FDS     &3.997    & 0.4137 & \bf{0.4205} & 0.4131 &0.2687 &0.2695 &0.2685\\
S-ListNet         & k=3  & ADS  &\bf{3.483}    & \bf{0.4184} & 0.4196 & \bf{0.4177} &\bf{0.2692} &\bf{0.2697} &\bf{0.2689}\\
\hline
S-ListNet         & k=4   & UDS  &6.161    & 0.4145 & 0.4226 & 0.4104 &0.2686 &0.2694 &0.2687\\
S-ListNet         & k=4   & FDS    &5.773    & 0.4145 & 0.4232 & 0.4150 &0.2690 &0.2695 &0.2686\\
S-ListNet         & k=4   & ADS &\bf{4.358}    & \bf{0.4149} & \bf{0.4247} & \bf{0.4164}& \bf{0.2692} & \bf{0.2700} & \bf{0.2689}\\
\hline
\end{tabular}
\end{center}
\caption{\label{tab:p@1OfMQ2008} Averaged training time (in seconds), P@1 and P@10 on training, validation (Val.) and test data with different Top-k methods. `C-ListNet' stands for conventional ListNet, `S-ListNet' stands for stochastic ListNet. }
\end{table*}

The averaged training time and the performance in precession are presented in Table~\ref{tab:p@1OfMQ2008}. For precession, both P@1
and P@10 results are reported, though we focus on P@1 since it is more concerned for applications such as QA.
Note that for stochastic ListNet, the optimal number of samples (document lists) has been selected according to the P@1 performance on the validate set.

From these results, it can be seen that the conventional Top-1 ListNet is rather fast, however
the Top-2 model is thousands of times slower. With $k > 2$, the training time
becomes prohibitive and so they are not listed in the Table. This is expected since the conventional ListNet considers
the full set of permutations which is a huge number with a large $k$. With the stochastic ListNet,
the training time is dramatically reduced. Even with a large $k$, the computation cost is still manageable,
because the computation is mostly determined by the number of object lists, rather than the value of $k$.
When comparing the three sampling methods, it can be found the convergence speed of the uniform distribution approach
is the slowest, probably due to the ineffective selection for relevant documents.
The adaptive distribution sampling is the fastest, probably attributed to the collaborative update
of the model and the distribution.

As for the P@1 performance, the stochastic ListNet method generally outperforms its non-stochastic
counterpart, particularly with the adaptive distribution
sampling. For example, the best P@1 results obtained on the test data with the stochastic Top-1
ListNet is $0.4127$, which outperforms the conventional
Top-1 ListNet ($0.4119$). This advantage of stochastic ListNet, as we argued, is largely attributed to its
capability of learning partial rank information with samples of partial sequences of the rank list.

Comparing the results with different $k$ values, it can be seen that a larger $k$ tends to offer better P@1
performance on the training set, with either the conventional ListNet or the stochastic ListNet.
For example, with the conventional ListNet, the results are $0.4101$ vs. $0.4119$ with the Top-1 and Top-2
models respectively. However, the performance
gap is rather marginal, and the advantage with the large $k$ does not propagate to the results on the test data
(as has been seen in Figure~\ref{fig:k1} and Figure~\ref{fig:k2}).
This indicates that for the conventional ListNet, the Top-1 model is not the only choice in the sense of
computation complexity, but also the best choice in the sense of P@1 performance.

For stochastic ListNet, the performance improves with $k$ increases. In contrast to the conventional ListNet,
this improvement propagates to the results on the test data. For example, with the adaptive distribution
sampling, the P@1 results on the training set are $0.4102$ vs. $0.4184$ with the Top-1 and Top-3
models respectively, and the results on the test data are $0.4121$ vs. $0.4177$ respectively.

\begin{figure}[!htb]
\centering
\epsfig{figure=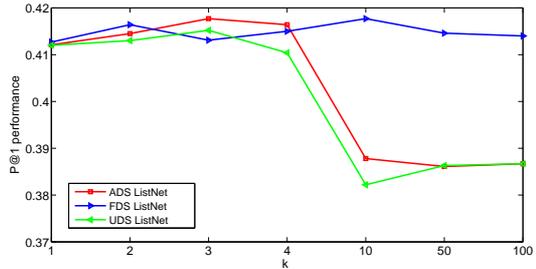,width=8cm}
\caption{The P@1 performance on the test data with the stochastic Top-k ListNet approach, where $k$ varies from $1$ to $100$.}
\label{fig:diffk}
\end{figure}

Nevertheless, the P@1 performance improvement with a large $k$ is rather marginal, and an over large $k$
simply reduces the performance. To make it clear, we vary the value of $k$ from $1$ to $100$ and plot the
P@1 results in Figure~\ref{fig:diffk}). It can be seen that larger $k$ ($> 4$) does not
offer any merit but causes performance instability, particularly with the adaptive sampling approach.
As we have discussed, with the stochastic ListNet,
partial rank information can be learned with simple Top-k models, even the Top-1 model.
This capability of partial rank learning with simple models reduces the necessity of employing
complex Top-k models. This is a highly valuable conclusion, and it suggests that a simple Top-1 or Top-2
model is sufficient for the ListNet method, if the stochastic method is applied. Considering
the trade-off between computation cost and model strength, we recommend stochastic Top-2 ListNet
which delivers better P@1 performance than the Top-1 model consistently, with sufficiently fast computing.
If more computation is affordable, stochastic Top-3 ListNet can be used to obtain better performance.

Finally, we highlight that the conclusions obtained from the P@1 results and the P@10 results perfectly match.
In fact, the P@10 results look more consistent between training and test data, and the advantage of the stochastic
approach seems more clear, particularly with the adaptive sampling. This is not surprising as the optimization
goal of ListNet is essentially to form a good rank that involves multiple candidates, and so P@10 is apt to measure
the superiority of a better rank approach.



\section{Discussion}
\label{sec:dis}

An interesting observation with the stochastic ListNet approach is that sampling more relevant documents improves performance.
This can be explained by the data imbalance between relevant and irrelevant documents, i.e., there are much
more irrelevant documents than relevant documents in the training data. This imbalance leads to biased models
that tend to classify all documents as irrelevant.
The re-sampling approach can be regarded as a way of balancing the two classes, and the fixed
and adaptive distribution sampling can be regarded as another way to achieve the goal. Note that in the fixed distribution sampling,
the distribution is solely dependent on the human-labeled scores. These scores are good measures of the \emph{rank} of relevance but not good
measures of the relevance itself. A possible way to solve this problem is to learn a scoring function that maps
human-labelled scores to more reasonable measures of document relevance, though we took a different way
that employs the network outputs as the relevance measures,
which is what the adaptive distribution sampling method does. Note that the network output is a natural
measure of document relevance,
so the adaptive distribution sampling works the best in our experiments.

Another related issue is the harsh labelling of the AM2008 dataset. In this dataset, documents are labelled
by only three values $\{0,1,2\}$, which is rather imprecise and the rank information is very limited. This harsh labeling is
another reason why the uniform distribution sampling does not work: by uniform distribution sampling, there is a large probability that the
sampled object lists involve documents that are all labelled by $0$. This leads to an inefficient learning. Another consequence of
the harsh labeling is that the power of complicated ranking models is largely constrained. For example, with the Top-k ($k > 1$)
ListNet model, many of the $k$ documents in a candidate list are labelled as the same score, resulting in limited rank information
for the Top-k model to learn. This is why Top-k models did not exhibit much superiority to the Top-1 model in our
experiments.  We argue that top-k models would provide more contributions with
more thorough labels (e.g., scores in real values). This is an ongoing research of our group.

Finally, we highlight that the stochastic approach is not limited to the ListNet model, but any model for listwise learning. It is well known that listwise
learning outperforms pairwise learning, due to it is capability of learning full ranks~\cite{liu2009learning}. However
learning full ranks requires unaffordable computation and so is infeasible in practice, even with the Top-k approximation. Our work demonstrated
that learning full ranks can be approximated by learning partial ranks, and a limited number samples of such partial ranks is sufficient
to convey the rank information. This stochastic learning is very fast, and even delivers better performance. It can be regarded as a general
framework that treats both the pair-wise learning and the full rank learning as two special cases. In fact, if the set of partial ranks involves
all the permutation classes, it reduces to the conventional listwise learning, and if the set of partial ranks involves all object pairs,
it resembles the pairwise learning. A wide range of listwise learning methods can benefit from the idea of stochastic learning provided
in this paper.

\section{Conclusion}
\label{sec:con}

This paper proposed a stochastic ListNet method to speed up the training of ListNet models and improve the ranking performance. The basic idea is to
approximate the full rank learning by learning a small number of partial ranks. Three sampling approaches were proposed to select the partial
ranks, and Top-k ListNet models with various complexity ($k$ values) were investigated.

Our preliminary results on the MQ2008 dataset confirmed that the stochastic ListNet approach can dramatically speeds up the model training,
and more interestingly, it can produce better ranking performance than the conventional ListNet. Especially, the adaptive distribution
sampling method delivered the best P@1 performance. An appealing observation is that the simple Top-2 model is very effective and more
complex Top-k models seem not very necessary, considering the trade-off between training complexity and model strength. This
observation, however, is purely based on the MQ2008 dataset. As have been discussed, more
detailed human labels may require more complex models, for which the stochastic method proposed in this paper is essential to conduct
the model training. For the future work, we plan to study Top-k ListNet models with other databases and apply the stochastic
learning approach to other listwise learning to rank methods.



\section*{Acknowledgments}

This research was supported by the National Science Foundation
of China (NSFC) under the project No. 61371136, and the
MESTDC PhD Foundation Project No. 20130002120011. It
was also supported by Sinovoice and Pachira.

\newpage

\bibliographystyle{acl}
\bibliography{references}

\end{document}